\documentclass[english,aps,prd,a4paper,groupedaddress,preprintnumbers,floatfix,nofootinbib,showpacs]{revtex4}

\usepackage{graphicx}
\usepackage{enumerate}
\usepackage{amsmath}
\usepackage{amssymb}
\usepackage{amsfonts}
\usepackage{xcolor}
\usepackage{url}
\usepackage{hyperref}
\hypersetup{colorlinks=true,linkcolor=redLinks,citecolor=greenLinks,
urlcolor=redLinks,
pdfborder={0 0 1}}
\hypersetup{
    bookmarks=true,         
    unicode=false,          
    pdftoolbar=true,        
    pdfmenubar=true,        
    pdffitwindow=false,     
    pdfstartview={FitH},    
    pdftitle={My title},    
    pdfauthor={Author},     
    pdfsubject={Subject},   
    pdfcreator={Creator},   
    pdfproducer={Producer}, 
    pdfkeywords={keyword1, key2, key3}, 
    pdfnewwindow=true,      
    colorlinks=false,       
    linkcolor=red,          
    citecolor=green,        
    filecolor=magenta,      
    urlcolor=cyan           
}
\colorlet{shadecolor}{gray!15}
\definecolor{greenLinks}{rgb}{0, 0.6, 0} 
\definecolor{blueLinks}{rgb}{0, 0, 0.6}
\definecolor{redLinks}{rgb}{0.6, 0, 0}
\definecolor{tempText}{rgb}{0.55, 0.10,0.67}
\definecolor{eprintLinks}{rgb}{0.4, 0.4, 0.4}
\definecolor{journalLinks}{rgb}{0.6, 0, 0}

\newcommand{\bi}{\begin{itemize}}
\newcommand{\ei}{\end{itemize}}

\newcommand{\Cinvestav}{Departamento de F\'{\i}sica, Centro de
  Investigaci{\'o}n y de Estudios Avanzados del IPN,\\ Apdo. Postal
  14-740, 07000 Ciudad de M\'exico, M\'exico.}
  
   \newcommand{\UNAM}{Instituto de F\'{\i}sica, Universidad Nacional Aut\'onoma de M\'exico,\\
   Apdo. Postal 20-364, 01000 Ciudad de M\'exico, M\'exico.}
  
\newcommand{\AddrAHEP}{AHEP Group, Institut de F\'{i}sica Corpuscular --
  C.S.I.C./Universitat de Val\`{e}ncia, \\Parc Cient\'{\i}fic de Paterna,
  C/Catedr\`{a}tic Jos\'e Beltran 2, E-46980 Paterna (Val\`{e}ncia), Spain.}

\begin{document}

\title{The weak mixing angle from low energy neutrino measurements: a global update}

\author{B. C. Ca\~nas$^1$}\email{bcorduz@fis.cinvestav.mx} 

\author{E. A. Garc\'es$^2$}\email{egarces@fisica.unam.mx} 

\author{O. G. Miranda$^1$}\email{OmarMiranda@fis.cinvestav.mx} 

\author{M. T\'ortola~$^3$}\email{mariam@ific.uv.es}

\author{J.W.F. Valle~$^3$}\email{valle@ific.uv.es, URL: http://astroparticles.es/} 

\affiliation{$^1$\Cinvestav}
\affiliation{$^2$\UNAM} 
\affiliation{$^3$\AddrAHEP} 


\begin{abstract}
  Taking into account recent theoretical and experimental inputs on
  reactor fluxes we reconsider the determination of the weak mixing
  angle from low energy experiments. We perform a global analysis to
  all available neutrino--electron scattering data from reactor
  antineutrino experiments, obtaining
  $\sin^2\theta_W = 0.252 \pm 0.030$.  We discuss the impact of the
  new theoretical prediction for the neutrino spectrum, the new
  measurement of the reactor antineutrino spectrum by the Daya Bay
  collaboration, as well as the effect of radiative corrections.  We
  also reanalyze the measurements of the $\nu_e - e$ cross section at
  accelerator experiments including radiative corrections. By
  combining reactor and accelerator data we obtain an improved
  determination for the weak mixing angle,
  $\sin^2\theta_W = 0.254 \pm 0.024$.
\end{abstract}


\maketitle


\section{Introduction}
\label{sec-intro}

The weak mixing angle is a fundamental structural parameter of the
Standard Model (SM) and it has been measured with great precision at
high energies~\cite{Agashe:2014kda}. At low energies, except for
atomic physics measurements~\cite{Erler:2004in}, its determination has
always been a difficult task, especially in neutrino experiments. On
the one hand reactor antineutrino scattering off electrons reported
results indicating a relatively large value of the weak mixing
angle~\cite{Barranco:2007ea,Deniz:2009mu}, without a strong
statistical significance. The importance of a new measurement of this
fundamental parameter in the low energy region has been stressed in
various works and several proposals have been discussed in this
direction~\cite{Conrad:2004gw,Agarwalla:2010ty,Garces:2011aa}.
On the other hand, the interaction of neutrinos with quarks at NuTev
energies gave measurements that appeared to be in disagreement with
the SM~\cite{Zeller:2001hh}, although a recent evaluation of the sea
quark contributions suggests agreement with the Standard Model
predictions~\cite{Ball:2009mk,Bentz:2009yy}.

Reactor neutrino experiments have provided a useful tool for measuring
antineutrino scattering off electrons over at least four
decades~\cite{Reines:1976pv} and more recent studies of this process
have led to improved
measurements~\cite{Deniz:2009mu,Amsler:1997pn,Derbin:1993wy,Vidyakin:1992nf}.
On the other hand one expects that new results may be reported in the
near future, for instance by the GEMMA experiment~\cite{Beda:2012zz}
which would help improving the current determinations
of the weak mixing angle.
Moreover, the MINERVA Collaboration has reported first
neutrino-electron elastic scattering measurement, providing an
important restriction on the relevant neutrino flux, useful to
future neutrino beams operating at multi-GeV
energies~\cite{Park:2015eqa}. 

Recently, a revaluation of the reactor antineutrino energy
spectrum~\cite{Mueller:2011nm,Huber:2011wv} has revived the issue of the possible
existence of a light sterile neutrino~\cite{Mention:2011rk}. In this
work, we study the impact of the new predicted reactor spectrum on the
evaluation of the weak mixing angle. In order to have a more
reliable result we will also include the effect of radiative
corrections upon the neutrino-electron scattering.  We will discuss
the interplay between the impact of the new reactor spectrum and the
radiative corrections, showing that the overall effect is a shift
towards the Standard Model prediction for the weak mixing
angle.
In order to reach this conclusion we analyze the available
neutrino-electron scattering data from the reactor experiments based
at the Kuo Sheng (TEXONO)~\cite{Deniz:2009mu,Wong:2006nx}, Bugey
(MUNU)~\cite{Daraktchieva:2005kn,Amsler:1997pn},
Rovno~\cite{Derbin:1993wy} and Krasnoyarsk~\cite{Vidyakin:1992nf}
sites
\footnote{ Notice that we are not including the
    pioneering Reines reactor experiment of
    Ref.~\cite{Reines:1976pv}. The lack of detailed publicly available
    information prevents an improved re-analysis of these data
    including radiative corrections in the cross section.}.
  We have also included accelerator experiments in our analysis, such
  as the measurements from LAMPF~\cite{Allen:1992qe} and
  LSND~\cite{Auerbach:2001wg}.  These are sensitive to the scattering
  of electron neutrinos with electrons, providing complementary
  information to reactor experiments. As a result we obtain a more
  precise determination for the weak mixing angle.

\section{The neutrino electron scattering measurement}
\label{sec:nu-e}

\subsection{Reactor experiments}

In order to perform an analysis of the reactor antineutrino data
scattering off electrons it will be necessary to compute the expected
number of events and compare it with the experimental results through
a statistical analysis. In this section we describe this procedure.
The number of events per energy bin for each experiment is, in general, given by
\begin{equation}
N_{i} = n_e\Delta t \int \int \int^{T'_{i+1}}_{T'_i}
                   \lambda(E_{\nu}) \frac{d\sigma(E_{\nu}, T) }{ dT}  R(T,T') dT' dT dE,
\label{eq:num-events}
\end{equation}
where $\lambda(E_{\nu})$ corresponds to the antineutrino spectrum and
$R(T,T')$ denotes the energy resolution function associated to the
detector. This function accounts for  possible differences between the observed electron recoil
energy $T'$ and its true value $T$, and it is parameterized as
\begin{equation}
\label{eq:resolution}
R(T,T')=\frac{1}{\sqrt{2\pi}\sigma}\exp\left\{-\frac{(T-T')^2}{2\sigma^2}\right\}, 
\end{equation}
with $\sigma=\sigma(T)=\sigma_0\sqrt{T/MeV}$. 
The differential weak cross section for antineutrino-electron scattering, at tree
level, can be expressed as
\begin{eqnarray}
\frac{d\sigma(E_{\nu}, T) }{ dT} &=& \frac{G_F^2 m_e} {2 \pi} \left[  (g_V - g_A)^2 + 
(g_V + g_A )^2\left(1-\frac{T}{E_\nu}\right)^2 - (g_V^2-g_A^2)\frac{m_eT}{E_\nu^2}   \right] ,
\label{cross-section}
\end{eqnarray} 
where $E_\nu$ is the incoming neutrino energy, $G_F$ is the Fermi
constant, $m_e$ is the electron mass and $T$ is the electron recoil
energy. At tree level, the coupling constants $g_V$ and $g_A$ are given by
\begin{equation}
g_V = \frac12 + 2\sin^2\theta_W \, ,~~~~~~~~ g_A = \frac12\, .
\end{equation}
Although the  tree level expression in Eq.~(\ref{cross-section})
is useful to show the main dependence on the weak mixing
angle, we will consider radiative corrections for the scattering of neutrino 
and antineutrino off electrons in all our calculations.
In particular, we will follow closely the prescriptions derived in
Refs.~\cite{Bahcall:1995mm,Sarantakos:1982bp} 
where the radiative corrections are included taking the value of the
weak mixing angle at the $Z$ peak in the $\overline{\text{MS}}$-scheme
and some energy-dependent functions to include the effect of running
with the scale.

We now proceed to  describe the procedure used to  re--evaluate the
weak mixing angle from reactor antineutrino data including radiative 
corrections.
First of all, in order to calculate the expected number of events for
antineutrino electron scattering off electrons as given by
Eq.~(\ref{eq:num-events}), we first need a model that predicts the
produced reactor antineutrino flux.
Recently, a new evaluation of the reactor antineutrino spectrum has
appeared in the literature~\cite{Mueller:2011nm,Huber:2011wv}, claiming that the
previous predictions were underestimating the total reactor
antineutrino flux by approximately 3\%~{\footnote{See
    Ref.~\cite{Hayes:2016qnu} for a recent review on antineutrino
    reactor spectra.}.
The new reactor antineutrino spectrum is parametrized by a combination of order five polynomial functions 
given by
\begin{equation}\label{eq:flux}
 \lambda(E_\nu) = \sum_{\ell} f_\ell \,  \lambda_{\ell}(E_{\nu}) = \sum_{\ell} f_\ell \, \exp\left[\sum_{k=1}^{6}\alpha_{k\ell}E_{\nu}^{k-1}\right] \, ,
\end{equation}
where $f_\ell$ is the fission fraction for the isotope $\ell$ $\equiv$ $^{235}$U,  $^{239}$Pu, $^{241}$Pu and
$^{238}$U, at the reactor under study.
The values of the coefficients $\alpha_{k\ell}$ for
  energies above $2$~MeV can be found at the original
  references~\cite{Mueller:2011nm,Huber:2011wv}. Here we will follow
  the prescriptions in Ref.~\cite{Mueller:2011nm}. For smaller
energies we use the reactor antineutrino spectrum given at
Ref.~\cite{Kopeikin:1997ve}.

After calculating the expected number of events at a given reactor
experiments, we perform a statistical analysis that, comparing the
predictions with the actually number of observed events, will give us
a determination of the weak mixing angle value. We start the
description of the $\chi^2$ analysis chosen with the treatment of the
systematic uncertainties for the antineutrino reactor spectrum.
In order to quantify the systematical uncertainties coming from the
reactor anti-neutrino flux, we follow the diagonalization method for
the covariance matrix discussed in~\cite{Huber:2004xh}.
We take into account the errors of the  $\alpha_{k\ell}$
coefficients, $\delta\alpha_{k\ell}$, and their corresponding correlation
matrix, $\rho^{\ell}_{k k'}$.  The covariance matrix in terms
of these quantities can be written as 
\begin{equation}\label{eq:covar}
V^\ell_{kk'} = \delta \alpha_{k\ell} \, \delta \alpha_{k'\ell} \, \rho^\ell_{kk'} \,.
\end{equation}
With this parameterization,  the systematic error in the number of
events associated to the reactor antineutrino flux is given by
\begin{equation}\label{eq:Nnutot}
(\delta N^\nu_\ell)^2 = \sum_{kk'} 
\frac{\partial N^\nu_\ell}{\partial \alpha_{k\ell}}
\frac{\partial N^\nu_\ell}{\partial \alpha_{k'\ell}} V^\ell_{kk'} .
\end{equation}
Note that for the numerical analysis it is better to work with the
diagonal form of the covariance matrix.  To this end, we introduce the 
new coefficients, $c_{k\ell}$, defined as
\begin{equation}\label{eq:rot}
\alpha_{k\ell} = \sum_{k'} \mathcal{O}^\ell_{k'k} \, c_{k'\ell} \,,
\end{equation}
where the rotation matrix $\mathcal{O}^{\ell}$ is given by
\begin{equation}
\mathcal{O}^{\ell} \, V^\ell \, (\mathcal{O}^\ell)^T
= \mbox{diag}\left[ (\delta c_{k\ell})^2 \right] \,.
\end{equation}
Thus, the new phenomenological parametrization of the flux in Eq.~(\ref{eq:flux}) can be rewritten as 
\begin{equation}
\lambda_\ell(E_\nu) = 
\exp\left[ \sum_{k=1}^{6} c_{k\ell} \, p^\ell_k(E_\nu) \right] \,,
\end{equation}
where $p^\ell_k(E_\nu)$ is a polynomial of $E_\nu$ given by
\begin{equation}\label{eq:poly}
p^\ell_k(E_\nu) = \sum_{k'=1}^{6} \mathcal{O}^\ell_{kk'}
E_\nu^{k'-1} \,.
\end{equation}

With all these ingredients we can now define the $\chi^2$ function we will use in our statistical analysis as
\begin{equation}
\chi^2_\mathrm{reactor}  = \sum_{ij} (N^{\rm theo}_i - N^{\rm exp}_i)
                 {\sigma^{-2}_{ij}}
                 (N^{\rm theo}_j - N^{\rm exp}_j) \, ,\
\end{equation}
where the expected number of events $N^{\rm theo}_{i}$ takes into account the contributions from all the
isotopes 
\begin{equation}
N_{i}^{theo}=N_{i}^{235}+N_{i}^{238}+N_{i}^{241}+N_{i}^{239},
\end{equation}
and $\sigma_{ij}^2$ is given as
\begin{equation}\label{eq:sigma2.2}
\sigma_{ij}^2=\Delta_{i}^2\delta_{ij} + \sum_\ell \delta N_i^\ell \delta N_j^\ell \, ,
\end{equation}
where $\Delta_i$ corresponds to the statistical uncertainty for the energy bin
$i$ and $\delta N_i^{\ell}$ is the contribution from the
isotope $\ell$ to the systematic error in the number of events at the same bin. This is
calculated as follows
\begin{equation}\label{eq:example}
\delta N_i^{\ell} = \sum_k \delta c_{k\ell} \, \frac{\partial N_i^{\ell}}{\partial c_{k\ell}} =
\sum_k \delta c_{k\ell} 
\int \int \int_{T'_i}^{T'_{i+1}}  \lambda_\ell(E_\nu) \, 
 p^\ell_k(E_\nu) \frac{d\sigma (E_\nu,T)}{dT}\, R(T,T') \, \, \mathrm{d}T' \, \mathrm{d}T \, \mathrm{d}E_\nu .
\end{equation}

Once we have set all the necessary tools for our analysis we will
describe in the next section the particular features of each reactor
experiment and we will present our results for the re--evaluation of
the weak mixing angle.

\subsection{Accelerator experiments}

Besides the reactor data, in this analysis we will include the
observation of neutrino scattering off electrons in accelerator
experiments.  In particular, we will use data from the
LAMPF~\cite{Allen:1992qe} and LSND~\cite{Auerbach:2001wg} experiments.
In this case, we will use as observable to fit the average cross
section at the experiment, given by
\begin{equation}
\sigma^{theo} = \int\int  
                   \lambda(E_{\nu}) \frac{d\sigma(E_\nu,T)}{dT} dT dE \, ,
\label{eq:num-ev-acc}
\end{equation}
where $\lambda(E_{\nu})$ is the electron neutrino flux coming from pion
decay~\cite{Allen:1992qe,Auerbach:2001wg} and  the
differential cross section for neutrino electron scattering is calculated as
\begin{eqnarray}
\frac{d\sigma(E_{\nu}, T) }{ dT} &=& \frac{G_F^2 m_e} {2 \pi} \left[  (g_V + g_A)^2 + 
(g_V - g_A )^2\left(1-\frac{T}{E_\nu}\right)^2 - (g_V^2-g_A^2)\frac{m_eT}{E_\nu^2}   \right] ,
\label{cross-section2}
\end{eqnarray} 
The statistical analysis of the neutrino accelerator data will be performed by using the following $\chi^2$ function
\begin{equation}\label{eq:chi2LSND}
 \chi^{2}_\mathrm{accel}
= \sum_{i=1}^{2 }\frac{(\sigma^{theo}_i(\sin^{2}\theta_{W})-\sigma^{exp}_i)^{2}}{(\Delta_i)^{2}},
\end{equation}
where the subindex $i=1,2$ stands for the LAMPF and LSND experiment,
respectively.  For the uncertainties we have included the statistical
and systematical errors on the reported cross section, added in
quadrature, as an uncorrelated error, $\Delta_i$.
To test our simulation, we have checked that the reported $1\sigma$
region for $\sin^2\theta_W$ is well reproduced with our simulation
once we ignore radiative corrections, as it was done at the original
references.

\section{Antineutrino-electron scattering at reactors}
  
\subsection{Summary of reactor data}
  
In this section we summarize the main  features of the
reactor antineutrino experiments relevant in our analysis.
\begin{itemize}
\item {\bf TEXONO.}
The latest experimental data from TEXONO were reported as a set of ten
energy bins ranging from 3 to 8 MeV in  electron kinetic recoil 
energy~\cite{Wong:2006nx}.
The fuel proportion at the reactor
($^{235}$U:$^{239}$Pu:$^{238}$U:$^{241}$Pu) was taken as
($0.55$:$0.32$:$0.07$:$0.06$) and the energy resolution function 
width equal to $\sigma=0.0325 \sqrt{T}$~\cite{Wong:2006nx}.
The data analysis of the TEXONO collaboration adopted the reactor
antineutrino spectrum reported in Ref.~\cite{Vogel:1989iv}.

\item{\bf MUNU.}  In the case of the antineutrino electron scattering
  measurements performed by the MUNU
  collaboration~\cite{Daraktchieva:2005kn}, the reactor fission
  fractions were reported to be ($0.54$:$0.33$:$0.07$:$0.06$).  The
  antineutrino spectrum originally considered for neutrino energies
  above $2$~MeV was the one reported in Ref.~\cite{Vogel:1989iv} as
  well, while the reactor antineutrino spectrum in
  Ref.~\cite{Kopeikin:1997ve} was adopted for lower energies.
The uncertainty in the  electron kinetic recoil  energy reconstruction was
parameterized with a resolution width given by $\sigma(T)=0.08
\text{T}^{0.7}$~\cite{Daraktchieva:2003dr}. 
Experimental measurements of the antineutrino-electron reaction were
presented in one single bin with electron recoil energy from 0.7 to 2
MeV, with a total of $1.07\pm 0.34$ counts per day (cpd) observed, in
agreement with the expectations of $1.02\pm 0.10$~cpd.

\item{\bf Rovno.}  The Rovno experiment~\cite{Derbin:1993wy}, measured
  the electron-antineutrino cross section at low recoil electron
  energies, in the range from 0.6 to 2 MeV.  For low energy antineutrinos
  they also used the theoretical prescription for the antineutrino
  spectrum reported in Ref.~\cite{Kopeikin:1997ve} while for energies
  above 2 MeV, the resulting antineutrino spectrum for $^{235}$U was
  taken from~\cite{Schreckenbach:1985ep}.
No information was given about the energy resolution function.
The Rovno experiment reported a measured cross section for neutrino
scattering by electrons equal to
$\sigma_{W}=(1.26\pm0.62) \times
10^{-44}$~cm$^2$/fission~\cite{Derbin:1993wy}.

\item{\bf Krasnoyarsk.}  This experiment observed the scattering of
  reactor antineutrinos with electrons for an electron recoil energy
  window in the range between 3.15 and 5.175 MeV, with a reported weak
  differential cross section given by
  $\sigma_{W}=(4.5\pm2.4)\times 10^{-46}$~cm$^2$/fission for
  $\sin^2\theta_{W}= 0.22$~\cite{Vidyakin:1992nf}.
  As in the case of the Rovno experiment, the initial neutrino flux
  coming from the $^{235}$U chain, as given in
  Ref.~\cite{Schreckenbach:1985ep}, was considered as the only 
  antineutrino source.

\end{itemize}

\subsection{Reactor data analysis with new antineutrino spectrum
  prediction}
\label{subsec:reac-mueller}

We show in Table~\ref{tab:tabla} a summary with the main details of
the reactor experiments described above. Besides the range of energy
explored at each experiment, we have also quoted the measured value
for the electron-antineutrino cross section, as well as the
value for the weak mixing angle, when reported.

\begin{widetext}
\begin{table}[!t]
\begin{center}
\begin{tabular}{l c c c c c c c } \hline \hline
Experiment               &  &  E$_\nu$(MeV)          &      &  T(MeV)      &  & Published cross-section    &     reported $\sin^2\theta_{W}$ \\ \hline \hline
TEXONO~\cite{Deniz:2009mu}    &     & $3.0-8.0$      &   &  $3\mbox{.}0-8\mbox{.}0$      & &  [1.08$\pm$0.21$\pm$0.16]$\cdot \sigma_{SM}$  &  $0\mbox{.}251\pm0\mbox{.}031\pm0\mbox{.}024$ \\ 
MUNU~\cite{Amsler:1997pn}       &  & $0\mbox{.}7-8\mbox{.}0$        & &  
$0\mbox{.}7-2\mbox{.}0$      & &  [1.07$\pm$0.34] events/day                        &    $\ldots$ \\
Rovno~\cite{Derbin:1993wy}       &  & $0\mbox{.}6-8\mbox{.}0$    &     &  $0\mbox{.}6-2\mbox{.}0$     &  &  [1.26$\pm$0.62]$\times10^{-44}$cm$^{2}$/fission&    $\ldots$ \\
Krasnoyarsk~\cite{Vidyakin:1992nf} & & $3\mbox{.}2-8\mbox{.}0$   &      &  $3\mbox{.}3-5\mbox{.}2$     &  &   [4.5$\pm$2.4]$\times10^{-46}$cm$^{2}$/fission &  $0\mbox{.}22_{-0\mbox{.}8}^{+0\mbox{.}7}$ \\ \hline \hline 
\end{tabular}
\caption{Summary of the measured $\bar{\nu}_{e}-e$ scattering cross
  sections and the corresponding $\sin^{2}\theta_{W}$ values obtained at the displayed reactor experiments.}
\label{tab:tabla}
\end{center}
\end{table}
\end{widetext}

In a previous global analysis of reactor and accelerator
neutrino-electron scattering data~\cite{Barranco:2007ea}, the weak
mixing angle value was found to be
\begin{equation}
  \label{eq:tw1}
\sin^2\theta_W=0.259\pm 0.025 \,.  
\end{equation}
Recently, an updated version of this analysis including the reactor
data from TEXONO has reported a slightly improved determination of the
weak mixing angle~\cite{Khan:2016uon}
\begin{equation}
  \label{eq:tw2}
\sin^2\theta_W=0.249\pm 0.020 \,.
\end{equation}
However, the role of radiative corrections in the weak cross section
has not been discussed in these references.
Both  results  lie  above  the  theoretical  predicted  value  in  the
$\overline{\text{MS}}$-scheme at the $Z$-peak~\cite{Agashe:2014kda}
\begin{equation}
  \label{eq:tw3}
\sin^2\theta_W=0.23126\pm 0.00005 \,.  
\end{equation}

In order to illustrate how sensitive is the weak mixing angle to the
presence of radiative corrections in the antineutrino-electron
scattering cross section and to the considered reactor antineutrino
spectrum, we present in Table~\ref{tab:tableReactors} the central
value of $\sin^2\theta_W$ obtained from each reactor antineutrino
experiment, under the following assumptions:
a) the original antineutrino spectrum considered in the original
analysis of the experimental collaboration without radiative
corrections,
b) the original spectrum  including radiative corrections,
c) the new reactor antineutrino spectrum without radiative
corrections, and
d)  the new reactor antineutrino spectrum  including radiative
corrections.  
In the first row we have reported the fit for the original spectrum
used in each experiment without including radiative corrections.  For
the TEXONO case, notice that the value obtained in the absence of
radiative corrections and with the original spectrum is in good
agreement with the value reported by the collaboration,
$\sin^2\theta_W = 0.251$~\cite{Deniz:2009mu}.
The next rows in the table show the separate effect of including
either the new antineutrino spectrum or the radiative corrections. Finally, the
last row shows our updated analysis with all the improvements included.
One can see from this table that the impact of the new analysis is
different for every experiment and some experiments give closer values
to the expected theoretical predictions than others.  For each
$\chi^2$ analysis we have taken into account the systematic error for
the antineutrino energy flux and the statistical errors.

\begin{table}[!t]
\begin{center}
\begin{tabular}{l c c c c c c } \hline \hline
& Mueller spectrum \quad      &  Radiative correc.    \quad &  TEXONO \quad & MUNU \quad & Rovno \quad & Krasnoyarsk   \\ \hline \hline
a) & -           &  -            &  $0.256$ & $0.241$ &  $0.220$ &  $0.220$  \\  
b) & -            &  $\checkmark$ &  $0.261$ & $0.248$ &  $0.226$ &  $0.224$  \\  
c) & $\checkmark$ &  -            &  $0.253$ & $0.237$ &  $0.228$ &  $0.231$  \\
d) & $\checkmark$ &  $\checkmark$ &  $0.258$ & $0.244$ &  $0.235$ &  $0.235$   \\ 
\hline \hline 
\end{tabular}
\caption{
  Weak mixing angle determinations obtained from reactor data using
  different assumptions for the antineutrino spectrum and radiative corrections, as indicated.
  }
\label{tab:tableReactors}
\end{center}
\end{table}

We have performed a combined statistical analysis using the data from
the four reactor experiments described above.  Our global
determination for the weak mixing angle is shown in
Fig.~\ref{fig:theta-reac}.
We also give in the plot the $\Delta\chi^2$ profiles obtained for each
reactor experiment individually.  As we can see, the most recent
TEXONO data play a dominant role in the combined analysis, although
the previous experiments shift the preferred value of $\sin^2\theta_W$
towards a slightly smaller central value:
\begin{equation}
\sin^2\theta_W = 0.252 \pm 0.030 .
\end{equation}

\begin{figure}[!t]
\includegraphics[width=0.6\textwidth]{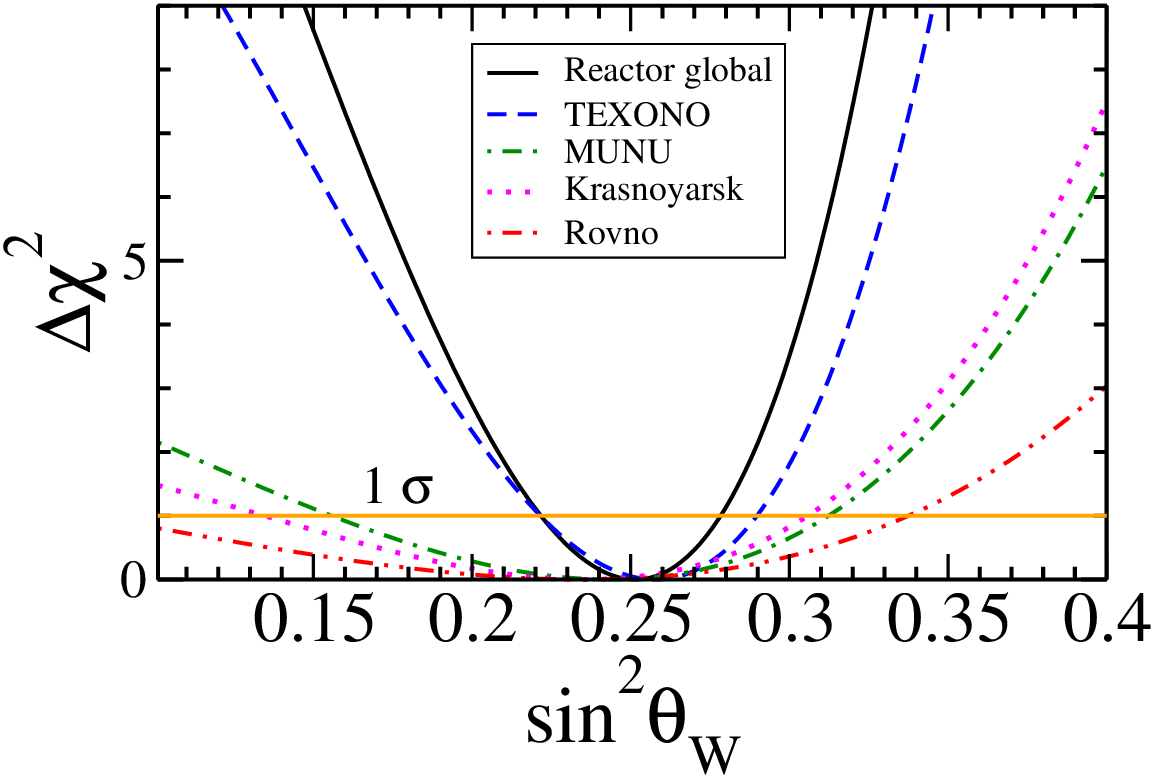}
\caption{\label{fig:theta-reac}
Determination of $\sin^2\theta_W$ from the combined analysis of reactor experiments (solid black line). 
The individual $\Delta\chi^2$ profiles obtained from each single experiment are also shown.}
\end{figure}

\subsection{Impact of the Daya Bay total reactor flux determination}

Recently, the Daya Bay collaboration has published results on the
measurement of the antineutrino spectrum using inverse beta
decay~\cite{An:2015nua}.  
There are indications that this measurement is not fully consistent
with the recent theoretical predictions for the antineutrino flux
produced at reactors~\cite{Mueller:2011nm,Huber:2011wv}.
Further theoretical developments and experimental measurements will be
required in order to settle this point. While this question is solved,
here we have estimated the impact of the recent Daya Bay reactor flux
measurement on the extraction of the weak mixing angle from reactor
data. As a first approximation, we correct the theoretical spectrum
predicted by Mueller et al.~\cite{Mueller:2011nm}
with the overall normalization factor $0.946$, which is the central
  value for the ratio of
  measured to predicted flux, as reported by the Daya Bay
  collaboration~\cite{An:2015nua,An:2016srz}. 
\begin{figure}[!tb]
\includegraphics[width=0.6\textwidth]{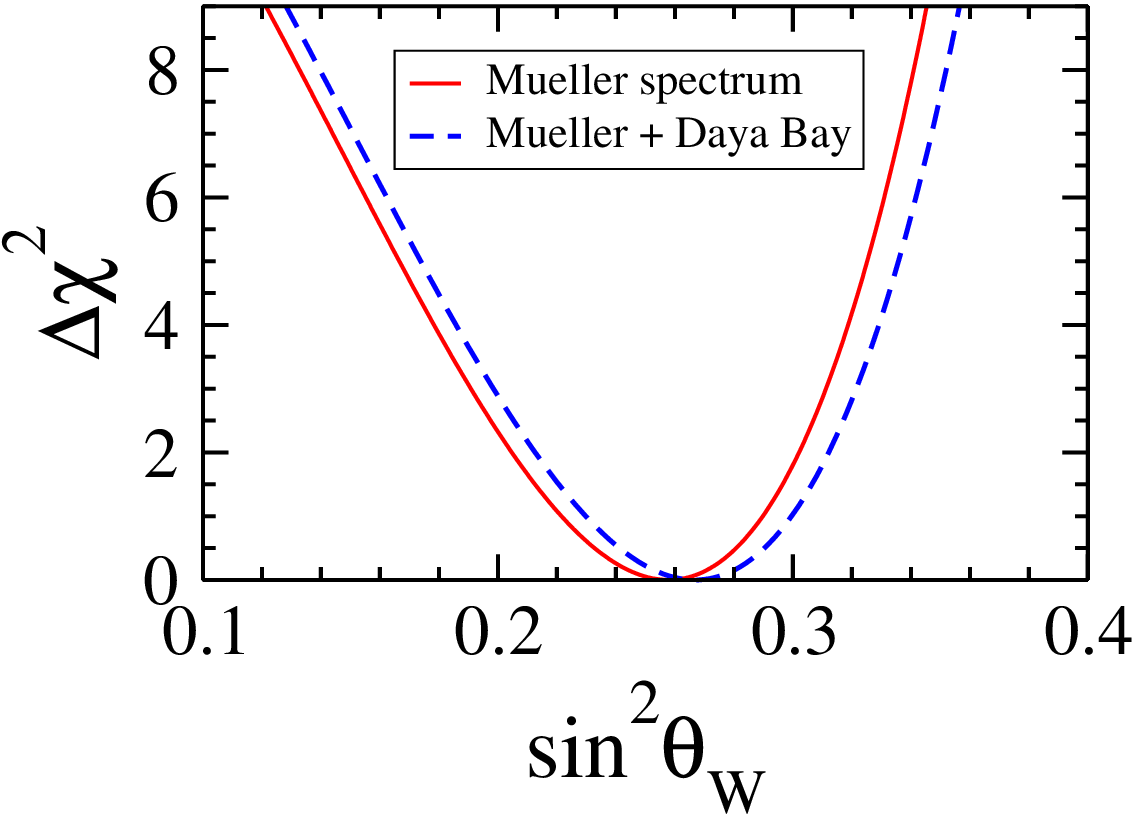}
\caption{\label{fig:dayabay} Determination of $\sin^2\theta_W$ from
  TEXONO data using the original Mueller et al. spectrum (solid red line)
  and the Mueller spectrum corrected by the Daya Bay measurement of
  the total reactor antineutrino flux (dashed blue line).  }
\end{figure}

\begin{figure}[!t]
\includegraphics[width=0.6\textwidth]{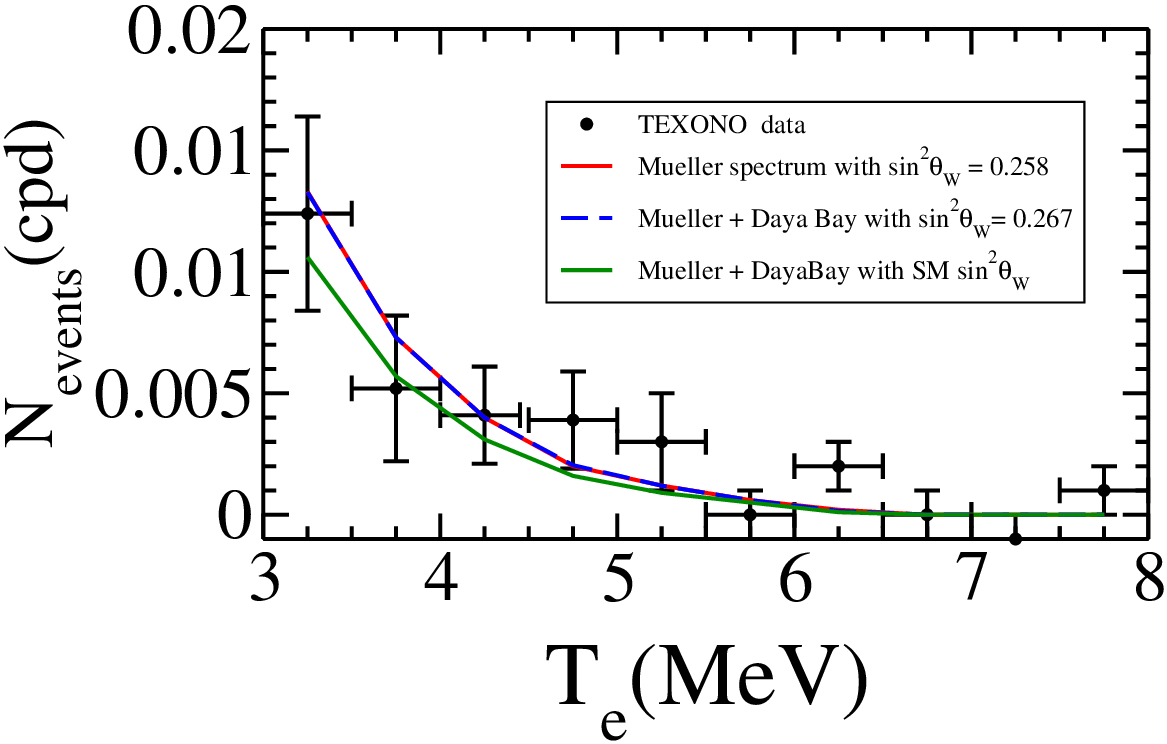}
\caption{\label{fig:3d} Expected event numbers in TEXONO using the
  Mueller et al. spectrum for the TEXONO $\sin^{2}\theta_{W}$ best fit
  value (solid red line). The blue dashed line corresponds to the best
  fit analysis obtained using the Mueller spectrum modified by the
  Daya Bay flux measurement.  The green solid line shows the
  prediction for the SM weak mixing angle, $\sin^2\theta_W =0.23126$.}
\end{figure}

The result of such an analysis for the TEXONO experiment is displayed
in Fig.~\ref{fig:dayabay}, where one can see that, if the Daya Bay
result is confirmed, the resulting value of weak mixing angle shifts
towards higher values.
\begin{equation}
\sin^2\theta_W = 0.267 \pm 0.033  \quad(\textrm{Mueller + DayaBay spectrum}).
\end{equation}

One sees that the TEXONO data correlates the flux normalization with
the value of the weak mixing angle, so that a decrease in the total
normalization prefers a higher value of $\sin^{2}\theta_W$.

In Fig.~\ref{fig:3d} we illustrate how the prediction of this analysis
compares with the experimental data from TEXONO. We plot the expected
number of counts per day in this experiment for three different
assumptions:
i) using the Mueller et al. spectrum~\cite{Mueller:2011nm} with
  the best fit value of the weak mixing angle obtained from the TEXONO
  data analysis, $\sin^2\theta_W = 0.258$, as in Section
  \ref{subsec:reac-mueller};
ii) the reactor antineutrino spectrum predicted by Mueller et al. with the
correction factor indicated by the Daya Bay measurements for the
obtained best fit value $\sin^2\theta_W=0.267$;
iii) the Mueller reactor antineutrino spectrum corrected by the Daya
Bay result for the SM prediction for the weak mixing angle at at the
$Z$-peak in the $\overline{\text{MS}}$ scheme:
$\sin^2\theta_W=0.23126$.
We can see from this figure how TEXONO data are in tension with the SM
prediction for the weak mixing angle, favoring higher values for
$\sin^2\theta_W$.  Further neutrino electron scattering measurements
will be necessary in order to have a better understanding both of the
neutrino reaction, as well as the reactor spectrum.

\section{Neutrino-electron scattering at accelerator experiments}

Besides studying electron scattering with electron antineutrinos
coming from reactors, in this work we have also analyzed the case of
electron neutrino scattering off electrons for two experiments that
used a spallation source.
In this case the electron neutrino flux came from pion decay and the
differential cross section was measured at the
LAMPF~\cite{Allen:1992qe} and LSND~\cite{Auerbach:2001wg} experiments.
Here we analyze the results on the neutrino-electron scattering cross
section reported by these experimental collaborations, given in
Table~\ref{tab:tab-acc}, using the procedure described in 
section~\ref{sec:nu-e}.
\begin{table}[!t]
\begin{center}
\begin{tabular}{l c c c} \hline \hline
Experiment   \quad      &  E$_\nu$(MeV)    \quad        \quad       & $\sigma^{\text{exp}}$ $\left[10^{-45}cm^{2}\right]$    &     reported $\sin^2\theta_{W}$ \\ \hline \hline
LAMPF\cite{Allen:1992qe}            &  7-50               &  [10.0$\pm$1.5$\pm$0.9]\,E$_{\nu}$ &  $0\mbox{.}249\pm0\mbox{.}063$\\  
LSND\cite{Auerbach:2001wg}       & 20-50               &  [10.1$\pm$1.1$\pm$1.0]\,E$_{\nu}$ &  $0\mbox{.}248\pm0\mbox{.}051$\\ \hline\hline 
\end{tabular}
\caption{Published $\nu_{e}-e$ scattering cross section and
  $\sin^{2}\theta_{W}$ measurements at accelerator experiments. The
  error combines the systematic and statistical uncertainties in both
  cases.  }
\label{tab:tab-acc}
\end{center}
\end{table}
After minimizing the $\chi^2$ function defined in
  Eq.~(\ref{eq:chi2LSND}) for both experiments, we obtained a new
  value for the weak mixing angle without radiative corrections:
\begin{equation}
 \sin^{2}\theta_W = 0.248 \pm 0.042 \,.
\end{equation}
The inclusion of radiative corrections to the neutrino-electron cross
section results in a somewhat higher value for the weak mixing angle:
\begin{equation}
\sin^{2}\theta_W = 0.261 \pm 0.042 \,.
\end{equation}
These results are given in Fig.~\ref{fig:4} and compared with the
results obtained from reactor experiments. They are also used to
obtain a global determination of the weak mixing angle from the
combination of reactor and accelerator data that will be discussed in
the next section. 

\section{Discussion and conclusions}

We have performed an updated analysis of the reactor and low--energy
accelerator neutrino experiments. In particular, we considered reactor
neutrino scattering off electrons. We have studied the impact of the
new reactor spectrum on the extracted value of the weak mixing angle.
The combined analysis shows an agreement with the theoretical
prediction, although more precise measurements in this energy range
would be highly desirable.
As illustrated in Table~\ref{tab:tableReactors}, using the new
spectrum prediction shifts the value for the weak mixing angle
differently for each reactor experiment. We show in Fig.~\ref{fig:4}
the expected value of $\sin^2\theta_W$ for a combined analysis of all
reactor experiments. We can see that in this case the inclusion of the
Mueller spectrum has a mild effect in the determination of
$\sin^2\theta_W$.

  We have also quantified the role of radiative corrections, both for
  neutrino and antineutrino scattering off electrons. The importance of
  radiative corrections can be seen in Fig.~\ref{fig:4} where we
  show the determination of the weak mixing angle with and without
  radiative corrections from reactor and accelerator 
  measurements of the (anti)neutrino-electron scattering cross section. 
  As one can see from the figure, the inclusion of radiative
  corrections increases the value of the weak mixing angle.
  From the combined analysis of all the experiments considered, we
  obtain an improved determination of the weak mixing angle
\begin{equation}
 \sin^{2}\theta_W = 0.254 \pm 0.024.
\end{equation}
This should be compared with other determinations at different energy
ranges, much more precise, as seen in Fig.~\ref{fig:5}.  
  In order to illustrate how this result compares with other low
  energy measurements, we can take, as a first approximation, the weak
  mixing angle at low energies as given as~\cite{Kumar:2013yoa}
\begin{equation}
\sin^2\theta_W(0)_{\overline{\text{MS}}} 
= \kappa(0)_{\overline{\text{MS}}}  \sin^2\theta_W(M_Z)_{\overline{\text{MS}}} 
\end{equation}
with $\kappa(0) = 1.03232$~\cite{Kumar:2013yoa}. This approach can
give an idea of the level of precision that has been reached by
neutrino electron scttering and it is shown in Fig.~\ref{fig:5}, where
we compile the most important measurements already
reported~\cite{Agashe:2014kda}.  
\begin{figure}[!t]
\includegraphics[width=0.6\textwidth]{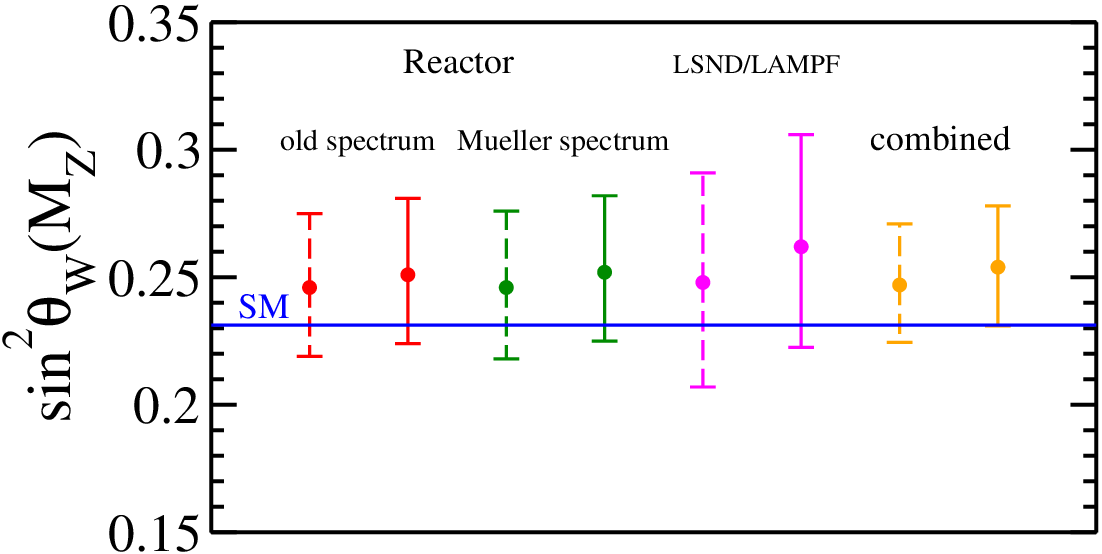}
\caption{\label{fig:4} Values of the weak mixing angle from the global
  analysis of reactor experiments using the original (old) or the
  Mueller reactor spectrum with (continuous error bars) and without
  (dashed error bars) radiative corrections. The combined result from
  the accelerator experiments LSND and LAMPF is shown for comparison,
  as well as the result of combining all the low-energy
  measurements. The horizontal line corresponds to the Standard Model
  prediction in the $\overline{\text{MS}}$ scheme.}
\end{figure}

\begin{figure}[!t]
\includegraphics[width=0.6\textwidth]{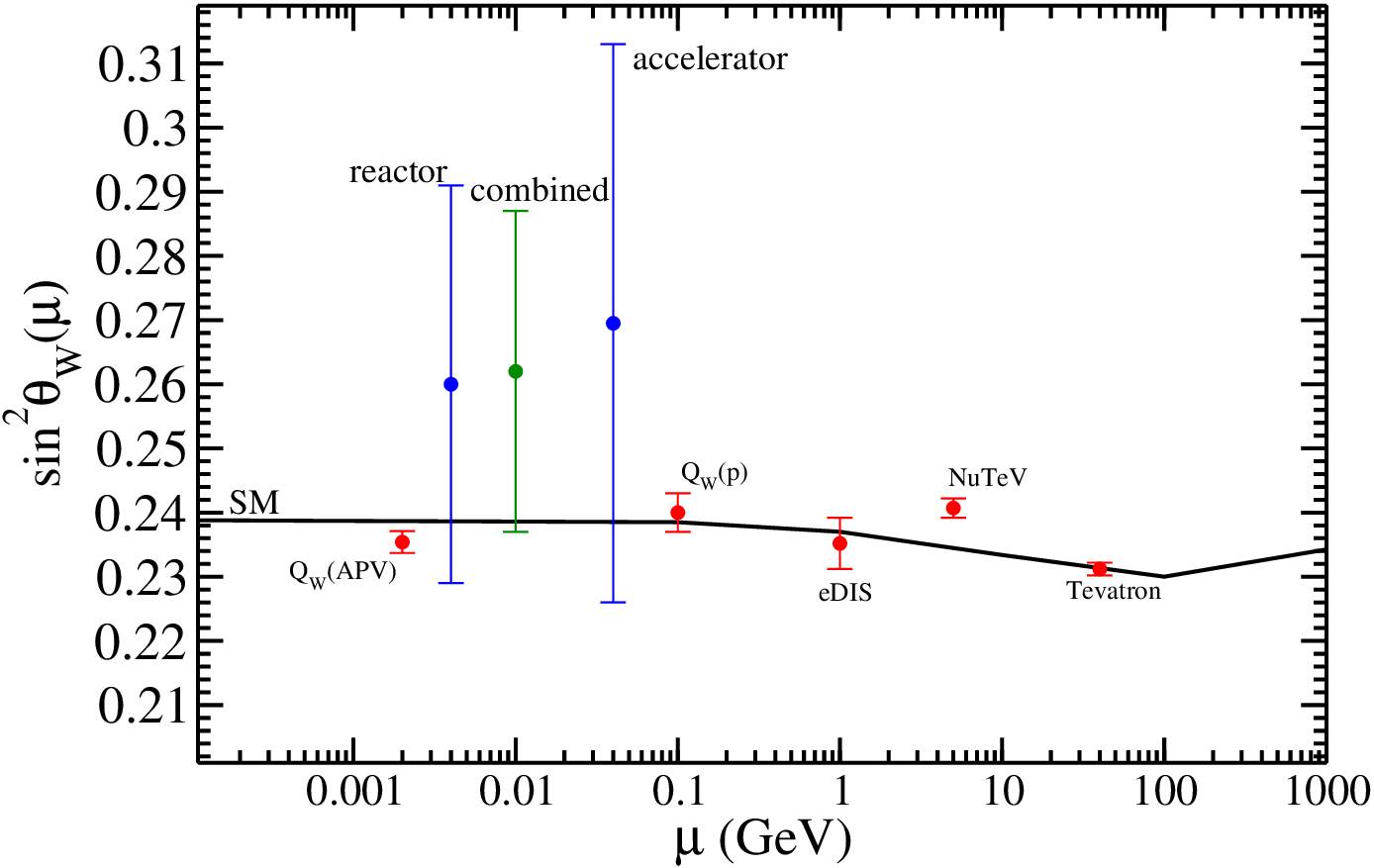}
\caption{\label{fig:5} 
Values of the weak mixing angle, in the
  $\overline{\text{MS}}$ scheme, from various experimental
  determinations, according to Ref.~\cite{Agashe:2014kda}. 
  For comparison, we extrapolate our results to the low-energy limit 
  as discussed in the text. 
  }
\end{figure}

Beyond the modest improvement we have obtained in our analysis, one
should stress the importance of further more refined experiments in
electron (anti)neutrino--electron scattering, so as to improve the low
energy determination of the weak mixing angle from neutrino
experiments.
Indeed, proposals such as GEMMA~\cite{Beda:2012zz} may provide better
Standard Model probes at low energies. On the other hand, they could
also open a window for important probes of neutrino properties and the
structure of the electroweak theory, since the experimental technique
itself seems not yet fully optimized~\cite{Conrad:2004gw}.
Moreover, they should provide improved reactor antineutrino flux
measurements.
Indeed, various proposals for improving neutrino electron scattering
measurements have been discussed in the literature, either using
reactor neutrinos or a proton beam~\cite{Agarwalla:2010ty}.
Another possibility would be the use of an upgraded version of the
Borexino detector~\cite{Bellini:2013lnn}, such as envisaged in the
framework of a LENA--like proposal~\cite{Wurm:2011zn}, either in
combination with solar neutrinos, or with an artificial neutrino
source~\cite{Garces:2011aa}.

\acknowledgments { Work supported by Spanish grants FPA2014-58183-P,
  Multidark CSD2009-00064, SEV-2014-0398 (MINECO), PROMETEOII/2014/084
  (Generalitat Valenciana), and by the CONACyT grant 166639.
  E.A.G. thanks postdoctoral CONACYT grant.  M.~T. is supported by a
  Ram\'{o}n y Cajal contract (MINECO).  M.~T. thanks the Physics
  Department of Cinvestav for hospitality during the final phase of
  this work and Santander Universidades for the funding received
  through {\it Programa Becas Iberoam\' erica, J\'ovenes Profesores e
    Investigadores}.}


\end{document}